\documentclass[sn-nature]{sn-jnl}

\usepackage{graphicx}%
\usepackage{multirow}%
\usepackage{amsmath,amssymb,amsfonts}%
\usepackage{amsthm}%
\usepackage{mathrsfs}%
\usepackage[title]{appendix}%
\usepackage{xcolor}%
\usepackage{textcomp}%
\usepackage{manyfoot}%
\usepackage{booktabs}%
\usepackage{algorithm}%
\usepackage{algorithmicx}%
\usepackage{algpseudocode}%
\usepackage{listings}%
\usepackage{tikz}
\usepackage{quantikz}
\usepackage{natbib}

\usepackage{caption}
\usepackage{subcaption}

\begin{document}

\title[Superposed parameterised quantum circuits]{Superposed parameterised quantum circuits}

\author{Viktoria~Patapovich}
\author{Maniraman~Periyasamy}
\author{Mo~Kordzanganeh}
\author{Alexey~Melnikov}

\affil{Terra Quantum AG, 9000 St.~Gallen, Switzerland}

\abstract{
Quantum machine learning has shown promise for high-dimensional data analysis, yet many existing approaches rely on linear unitary operations and shared trainable parameters across outputs. These constraints limit expressivity and scalability relative to the multi-layered, non-linear architectures of classical deep networks. We introduce superposed parameterised quantum circuits to overcome these limitations. By combining flip-flop quantum random-access memory with repeat-until-success protocols, a superposed parameterised quantum circuit embeds an exponential number of parameterised sub-models in a single circuit and induces polynomial activation functions through amplitude transformations and post-selection. We provide an analytic description of the architecture, showing how multiple parameter sets are trained in parallel while non-linear amplitude transformations broaden representational power beyond conventional quantum kernels. Numerical experiments underscore these advantages: on a 1D step-function regression a two-qubit superposed parameterised quantum circuit cuts the mean-squared error by three orders of magnitude versus a parameter-matched variational baseline; on a 2D star-shaped two-dimensional classification task, introducing a quadratic activation lifts accuracy to 81.4\% and reduces run-to-run variance three-fold. These results position superposed parameterised quantum circuits as a hardware-efficient route toward deeper, more versatile parameterised quantum circuits capable of learning complex decision boundaries.}




\maketitle

\section{Introduction}\label{sec:intro}

The field of quantum machine learning (QML) formally began with the introduction of the variational quantum eigensolver (VQE) to find the ground state of the $\mathrm{HeH}^+$ molecule~\cite{Peruzzo2014Variational}. It involved applying a series of parameterised group rotations to a quantum circuit to prepare a quantum state and measuring this state in the Hamiltonian of the molecule to obtain the corresponding expected state energy~\cite{Tilly2022Variational}. The parameters were then tuned to minimise this energy, and the parameters corresponding to the lowest possible energy presented a reasonable approximation to the system's ground state. Subsequently, generalised parameterised quantum circuits (PQC) were extended from VQEs to machine learning (ML), where parameterised gates served two purposes: encoding classical data onto the quantum system and as trainable parameters~\cite{Melnikov2023Quantum,Benedetti2019Parameterized,Schuld2019Quantum}. The emergence of QML models sparked an endeavour to compare the PQC quantum models with their classical counterparts. Theoretical and practical comparisons shed light on the universality~\cite{Schuld2021Effect}, generalisation ability~\cite{Abbas2021Power,Caro2022Generalization}, and practical utility~\cite{Kordzanganeh2021Quantum,sagingalieva2025hybrid,kurkin2025forecasting} of these models.

A crucial result of this comparison was the recognition that quantum models used for supervised learning are kernel methods~\cite{Schuld2019Quantum,Havlicek2019Supervised}. Kernel methods use static feature maps to encode data samples onto a high-dimensional vector space and separate them using a trainable high-dimensional hyperplane~\cite{Schuld2018Supervised}. Similarly, PQCs encode the classical data onto a large Hilbert space and use the variational gates to train the hyperplane. To achieve a quantum advantage in machine learning, the equivalence of quantum models and kernel methods has profound consequences in finding classically-intractable kernels~\cite{Havlicek2019Supervised} or datasets where quantum kernels could prove beneficial~\cite{Huang2021Power}. However, kernel methods are limited to linear manipulations and fail to exhibit the success of deep learning models that are often attributed to their non-linear intermediate representations~\cite{Goodfellow2016Deep}.

Quantum models are also affected by the problem of barren plateaus~\cite{McClean2018Barren}: the gradients of a deep, randomly-initialised quantum circuit decay exponentially with the increasing number of qubits, except for some architectures~\cite{Zhao2021Analyzing}. Moreover, \cite{You2021Exponentially} showed that shallow quantum models with many qubits possess an exponentially-increasing number of sub-optimal local minima. Furthermore, \cite{Kordzanganeh2023Benchmarking} offered a practical runtime benchmark of increasing-width quantum models and showed that current quantum computers could provide a practical runtime speedup for simulating quantum circuits with over 30 qubits but with low state fidelity. These limitations mean that quantum models must be wide, deep, and able to surpass barren plateaus. \cite{Schuld2022Quantum} argued for a shift of focus from quantum advantage to three areas: quantum perceptron design, creating new links between classical and quantum machine learning theories, and upgrading the practicality of the infrastructure of QML. This work aims to contribute to the first two of these points.

In light of these challenges, there has been growing interest in creating novel architectures that combine advanced data encodings, multiple independent parameter sets, and controlled forms of non-linearity \cite{Schuld2022Quantum}. The present work introduces Superposed Parameterised Quantum Circuits (SPQCs) as one potential solution. Our construction uses Flip-Flop Quantum Random Access Memory (FFQRAM) to address multiple distinct parameter settings in a single circuit, thus allowing many sub-models or outputs to be realized in superposition \cite{Park2019Circuit}. At the same time, we exploit repeat-until-success (RUS) protocols to induce polynomial activations that emulate non-linear transformations on the state amplitudes \cite{Koppe2023Amplitude}. 

By integrating these elements, SPQCs aim to provide a multi-layered framework whose representational power parallels that of classical deep networks while still retaining the exponential-dimensional feature spaces inherent to quantum systems. We empirically demonstrate on both one-dimensional regression and two-dimensional classification datasets that SPQCs can learn complex boundaries more effectively than basic PQCs. Moreover, increasing the number of ancillary qubits or introducing polynomial activations leads to consistent improvements in training outcomes.

This paper is organized as follows. Section~\ref{sec:background} reviews related work on kernel-based quantum models, non-linearity in QML, and ensemble methods. In Section~\ref{sec:architecture}, we formally present the SPQC architecture and describe how FFQRAM and RUS techniques are combined. We then detail our experimental results in Section~\ref{sec:experiments}, highlighting the advantages of SPQCs over simpler quantum models. Finally, Section~\ref{sec:discussion} discusses practical limitations, and Section~\ref{sec:conclusion} concludes with an outlook on future directions for SPQCs and related QML architectures.

\section{Related Works and Background}\label{sec:background}
QML builds upon concepts from both the classical domain — such as multi-layered perceptrons (MLPs), kernel methods, and ensemble strategies — and the quantum domain, which introduces unitary evolution, superposition, and entanglement. This section provides a survey of the foundational ideas that underpin our proposed SPQC architecture. We first describe how classical neural networks leverage non-linearity to enhance expressivity, then review PQCs as the central paradigm of quantum model design. Finally, we outline two key techniques — RUS and FFQRAM — that address the limitations of purely linear quantum models and enable the parallel handling of multiple parameter sets in a single quantum run.

\subsection{Classical Neural Networks and Non-linearity}\label{subsec:classical-mlp}
Neural networks (NNs) have become a cornerstone of modern machine learning, achieving state-of-the-art results in image recognition \cite{Krizhevsky2012ImageNet,He2015Deep,Simonyan2015Very}, natural-language processing \cite{Vaswani2017Attention,Devlin2019Bert,Brown2020Language}, and speech synthesis \cite{Oord2016Wavenet,Shen2018Natural}. At the heart of every NN lie interconnected layers of simple computational units, or neurons \cite{rosenblatt1958perceptron}. A classical feed-forward network, commonly referred to as a multi-layered perceptron (MLP), comprises an input layer, one or more hidden layers, and an output layer \cite{Goodfellow2016Deep}. Each neuron in a hidden layer computes a weighted sum of inputs from the preceding layer and applies a non-linear activation function to the result:
\begin{equation*}
    y_i = \sigma\Biggl(\sum_{j} w_{ij} x_j + b_i\Biggr),
\end{equation*}
where \(x_j\) denotes the inputs from the previous layer, \(w_{ij}\) are trainable weights, \(b_i\) is a bias term, and \(\sigma(\cdot)\) is a non-linear activation function (e.g., sigmoid, ReLU, or tanh). The parameters \(w_{ij}\) and \(b_i\) across all layers of the network are tuned via numerical optimization to minimize a chosen loss function \cite{werbos1974beyond,rumelhart1986learning}.

Nonlinear activation functions endow neural networks with the capacity to model highly complex, nonlinear relationships in data. In their absence, even an arbitrarily deep network collapses to a single affine transformation, limiting the model to linear decision boundaries and precluding the capture of higher-order correlations \cite{Hornik1989Multilayer}. Indeed, classical universal-approximation theorems show that a feed-forward network with just one hidden layer and a non-constant bounded activation can approximate every continuous function on a compact set to arbitrary precision, given sufficiently many hidden neurons \cite{Hornik1989Multilayer,Cybenko2012Approximation}. More recent depth-separation results demonstrate that increasing depth, together with suitable nonlinearities, yields exponential gains in representational efficiency over shallow counterparts \cite{Telgarsky2016Benefits,Mhaskar2016DeepVsShallow}. A complementary line of work interprets deep networks as hierarchical kernel machines, attributing their inductive power to the repeated composition of nonlinear maps \cite{Poggio2017TheoryDeep}. Together, these theoretical insights help explain the empirical success of modern deep-learning systems across vision, language, and speech.

In practice, incorporating non-linear activations not only endows MLPs with a rich hypothesis space but also facilitates distributed representations within the layers. Such representations enable different neurons to specialise in distinct features of the data, leading to better generalization \cite{hinton1986learnin,bengio2013representation}. The recent surge in performance of deep learning can be largely attributed to both algorithmic advances (e.g., improved training methods and regularisation) and architectural innovations that exploit diverse nonlinearities
\cite{kingma2014adam,Ioffe2015BatchNorm,srivastava2014dropout,nair2010relu,he2016resnet,hendrycks2016gelu,ramachandran2017swish,clevert2015fast,misra2020mish}.

\subsection{Parameterised Quantum Circuits}\label{subsec:quantum-mlp}

QML seeks to exploit quantum processors to achieve performance or expressivity advantages~\cite{kordzanganeh2023exponentially} in tasks such as classification, regression and generative modelling \cite{Schuld2014Quest,Biamonte2017QuantumML,Cerezo2021VQA}. 
Among the various QML paradigms, parameterised quantum circuits (PQCs) — also known as variational quantum circuits or quantum neural networks — have become the tool of choice for near-term devices \cite{Benedetti2019Parameterized,Cong2019QCNN,Abbas2021Power}. 
A PQC realises a trainable function by embedding both data-dependent angles and continuous parameters into a quantum circuit and extracting classical information via measurement \cite{Mitarai2018QCL,Schuld2019Quantum,Havlicek2019Supervised}.

A typical supervised-learning PQC comprises three conceptual blocks:
\begin{enumerate}
  \item \textbf{Data encoding.} The input vector $\mathbf{x}$ is mapped to a quantum state with an encoding unitary $S(\mathbf{x})$. Encoding strategies range from single-qubit rotations to repeated \emph{data re-uploading}, which interleaves data injection and variational gates to boost expressivity \cite{Havlicek2019Supervised,PerezSalinas2020DataReupload}.
  \item \textbf{Variational layer.} A sequence of parameterised single-qubit rotations and entangling gates implements a tunable unitary $U(\boldsymbol{\theta})$, producing the state $|\psi_{\mathbf{x},\boldsymbol{\theta}}\rangle$ \cite{Benedetti2019Parameterized,Cong2019QCNN}.
  \item \textbf{Measurement.} Measuring an observable $M$ yields an expectation value $\langle\psi_{\mathbf{x},\boldsymbol{\theta}}|M|\psi_{\mathbf{x},\boldsymbol{\theta}}\rangle$, which serves as the model output \cite{Mitarai2018QCL}.
\end{enumerate}

The parameters $\boldsymbol{\theta}$ are optimised in a hybrid quantum–classical loop. 
Gradients can be obtained with the parameter-shift rule \cite{Crooks2019ParamShift}, the quantum natural gradient \cite{Stokes2020QNG}, or doubly stochastic gradient-descent schemes tailored to noisy expectation values \cite{Sweke2020SGD}. 
Recent information-geometric analyses show that well-designed PQCs can achieve a higher effective dimension — and therefore potentially stronger generalisation — than comparable classical models \cite{Abbas2021Power}.

Although the state space of \(n\) qubits is exponentially large, a
PQC followed by a fixed measurement
implements the map
\(
f_{\boldsymbol\theta}(\mathbf{x})
   =\langle\psi_{\mathbf{x},\boldsymbol\theta}|M|
     \psi_{\mathbf{x},\boldsymbol\theta}\rangle,
\)
which is strictly \emph{linear} in the data-encoded density operator.
Accordingly, standard PQCs can be recast as quantum kernel machines
that learn only linear decision boundaries in a (possibly vast) feature
Hilbert space \cite{Havlicek2019Supervised,Schuld2021Supervised}.
Linear models are already useful, yet classical deep networks owe
their universal approximation power to explicit non-linear activations.
An analogous resource in QML could enlarge hypothesis classes, boost
effective dimension and improve sample efficiency
\cite{Abbas2021Power,Holmes2023Nonlinear}.
Achieving controllable non-linearity on quantum hardware is
non-trivial, because unitary evolution alone cannot realise such
operations.  Recent proposals rely on adaptive circuits with
mid-circuit measurements and classical feedback
\cite{DeCross2023QubitReuse}, ancilla-assisted post-selection
\cite{Holmes2023Nonlinear}, or RUS primitives that iterate a compact sub-circuit until a heralded flag indicates success
\cite{Paetznick2013RUS,Guerreschi2019RUS}.
RUS circuits are particularly attractive for near-term devices because
they require only a small number of ancilla qubits, preserve coherence
between attempts, and have already been used to implement smooth,
trainable activation functions on superconducting processors
\cite{Moreira2023RUS}.
The next subsection provides the necessary background on RUS methods
and explains how they serve as the non-linear building blocks in our
enhanced SPQC architecture.

\subsection{Repeat-Until-Success Circuits}
\label{subsec:RUS-background}
One of the few circuit--level mechanisms capable of introducing such non--linear behaviour is the RUS paradigm originally introduced by \cite{Paetznick2013RUS}. An RUS construction appends one or more ancilla qubits to the data register, applies a carefully chosen sequence of entangling gates, and finally measures the ancillas. If the measurement returns a ''designated success'' pattern, the data register has undergone a target unitary $U_{\text{t}}$; otherwise a ''reversible failure'' unitary $U_{\text{f}}$ is applied and the whole routine is repeated. Because the success probability generally depends on the amplitudes of the data state, the overall, postselected transformation of the register is non--linear in those amplitudes: amplitudes that are twice as large do not simply double the success probability, but are renormalised by a state--dependent factor.

Subsequent works have refined the original idea. \cite{Bocharov2015RUS} and \cite{Wiebe2014RUSCircuit} demonstrated that families of RUS circuits can synthesise arbitrary real analytic functions of a single rotation angle with polylogarithmic overhead, while the so--called gearbox circuits of \cite{Koppe2023Amplitude} use two RUS layers to realise efficient amplitude‐dependent step and polynomial functions. These advances establish RUS as a versatile toolbox for embedding activation--like non--linearities directly in coherent quantum computations.

Taken together, these advances establish RUS circuits as a hardware-efficient, noise-robust, and functionally expressive building block for next-generation quantum machine-learning models. In the remainder of this paper we integrate RUS modules into an ensemble of shallow PQCs, thereby lifting the linearity inherent in purely unitary models while remaining compatible with gate-based devices available today.

\subsection{Flip–Flop Quantum Random Access Memory}\label{sec:theory:ff-qram}

Quantum Random Access Memory (QRAM) provides an \(n\)-qubit address register with coherent access to \(2^{n}\) classical data items \cite{Giovannetti2008PRL,Giovannetti2008PRA}. 
The availability of such a data oracle underpins several QML proposals, including kernel-based classification \cite{Rebentrost2014QSVM}. 

FFQRAM is a circuit-level realisation of this concept \cite{Park2019Circuit}. 
A register of $m$ ancillary address qubits is first placed in the uniform superposition  
\[
\frac{1}{\sqrt{L}}\sum_{k=0}^{L-1}\lvert k\rangle_{a},
\qquad L=2^{m},
\]  
after which a cascade of uniformly controlled rotations writes classical parameters \(\{\boldsymbol{\theta}^{(k)}\}_{k=0}^{L-1}\) onto an \(n\)-qubit data register. 
The resulting entangled memory state is 
\[
\lvert\Phi\rangle
  =\frac{1}{\sqrt{L}}\sum_{k=0}^{L-1}
    \lvert k\rangle_{a}\otimes\lvert\psi(\boldsymbol{\theta}^{(k)})\rangle_{d}.
\]  
Preparing this state requires \(O(nL)\) one- and two-qubit gates; the non-Clifford \(T\)-depth scales only logarithmically in \(n\) when RUS synthesis and ancilla recycling are used \cite{Bocharov2015RUS,Guerreschi2019RUS}.  

In the architecture presented here, the address register indexes an exponential family of variational layers, and the controlled rotations encode the corresponding parameter sets \(\boldsymbol{\theta}^{(k)}\) on the computational qubits, following the scheme of \cite{Yuan2023Optimal}. 
This layer-superposition technique allows \(L\) distinct parametrisations to be executed in parallel with just \(m=\log_{2}L\) additional qubits and sub-linear circuit depth, a key advantage for near-term, gate-based hardware.

\section{Architecture of SPQC}
\label{sec:architecture}

Conventional PQCs implement a linear kernel map between a data‐encoding unitary and a fixed measurement \cite{Schuld2019Quantum,Havlicek2019Supervised}. 
The SPQC introduced here augments this model with FFQRAM parallelisation \cite{Park2019Circuit} and RUS non-linear blocks \cite{Bocharov2015RUS,Guerreschi2019RUS}.

Each SPQC layer is synthesised in three steps. 
First, distributed PQCs supply prediction independence: $L$ disjoint parameter sets $\{\boldsymbol{\theta}^{(j)}\}_{j=0}^{L-1}$ are assigned to otherwise identical baseline PQCs, avoiding the shared-weights bottleneck of post-selected observables. 
Second, parallelisation with FFQRAM loads all $L$ sub-models onto a single data register using only $m=\log_{2}L$ ancillas. 
Third, an RUS gearbox realises amplitude functions $p_j\!\mapsto\!f(p_j)$; e.g. with two copies, $f(p)=p^{2}$, yielding a quadratic activation \cite{Bocharov2015RUS}. 
Successive layers repeat the procedure, so the SPQC stores the previous layer’s outputs coherently in the statevector amplitudes, exactly as fully connected matrices propagate activations in a classical MLP.

\subsection*{Distributed PQCs}
Suppose we wish to obtain multiple predictions $p_j$ from a quantum model. A common strategy is to keep a single parameterised circuit fixed and vary only the measurement observable $M$ — typically via post-selection — to extract different scalar outputs. Because every prediction traverses the same trainable layers, the resulting values are inevitably correlated \cite{Schuld2019Quantum}. 

\begin{figure}[htbp]
    \centering
    \includegraphics[width=\textwidth]{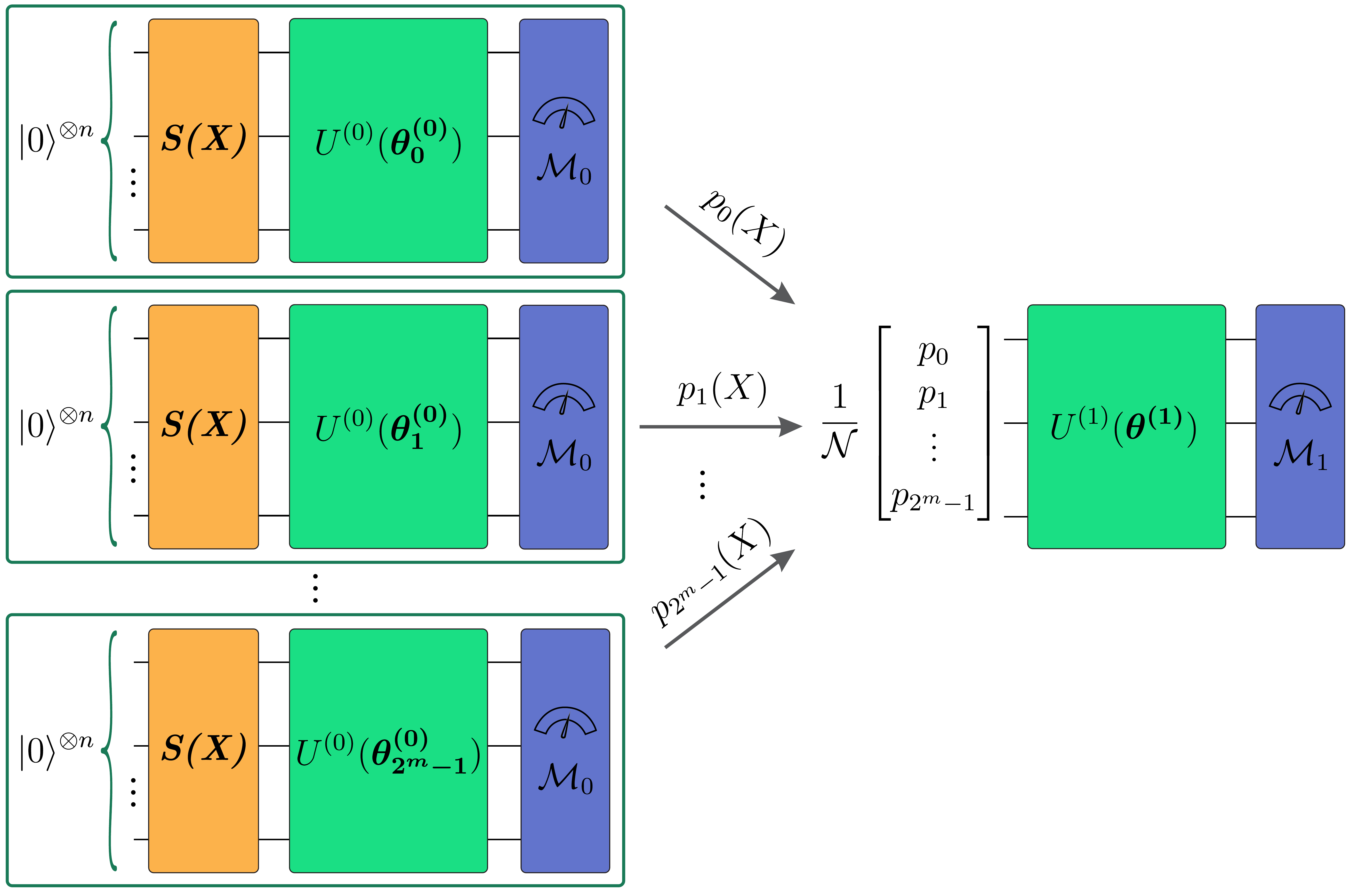}
    \caption{Resource-intensive replica strategy: $L$ copies of a PQC with parameters $\boldsymbol{\theta}^{(j)}$ produce independent outputs $p_{j}$ that feed a downstream model.}
    \label{fig:bagging}
\end{figure}

A direct route to statistical independence is to instantiate $L$ replicas of the baseline PQC, each equipped with its own parameter set $\boldsymbol{\theta}^{(j)}$. 
These copies act in parallel on the same input, producing the vector $\mathbf{p}=(p_0,\ldots,p_{L-1})$, which can then feed a downstream learner. 
The procedure mirrors classical bagging ensembles \cite{Breiman1996Bagging} but scales qubit count and gate depth linearly with $L$ (Fig.~\ref{fig:bagging}); on near-term hardware such overhead is prohibitive. FFQRAM mitigates this cost by loading all $L$ parameter sets into superposition with only $m=\log_2 L$ ancillary address qubits. 
We therefore adopt FFQRAM-based parallelisation throughout the remainder of this work.

\subsection*{Parallelisation with FFQRAM} FFQRAM reduces the resource overhead from linear to logarithmic in the number of sub-models~\cite{Park2019Circuit}. We append $m=\log_{2}L$ ancillary address qubits to the original $n$-qubit data register and prepare them in the uniform superposition $\frac{1}{\sqrt{L}}\sum_{j=0}^{L-1}\lvert j\rangle_{a}$ using a layer of Hadamard gates. A cascade of uniformly controlled rotations then writes the parameterised unitary $U\!\bigl(\boldsymbol{\theta}^{(j)}\bigr)$ onto the data qubits, conditioned on the address value, yielding \begin{equation} \lvert\Phi\rangle =\frac{1}{\sqrt{L}}\sum_{j=0}^{L-1} \lvert j\rangle_{a}\otimes U\!\bigl(\boldsymbol{\theta}^{(j)}\bigr) S(\mathbf{x})\lvert 0\rangle^{\otimes n}, \label{eq:qram_state} \end{equation} with $T$-depth scaling as $O(\log n)$ when RUS synthesis is employed~\cite{Guerreschi2019RUS}. Measuring the data register in the projector $\mathcal{M}=|0\rangle\!\langle0|^{\otimes n}$ collapses the address qubits to amplitudes proportional to the prediction vector $p_j$. 

\begin{figure}[htbp]
  \centering
  \includegraphics[width=\textwidth]{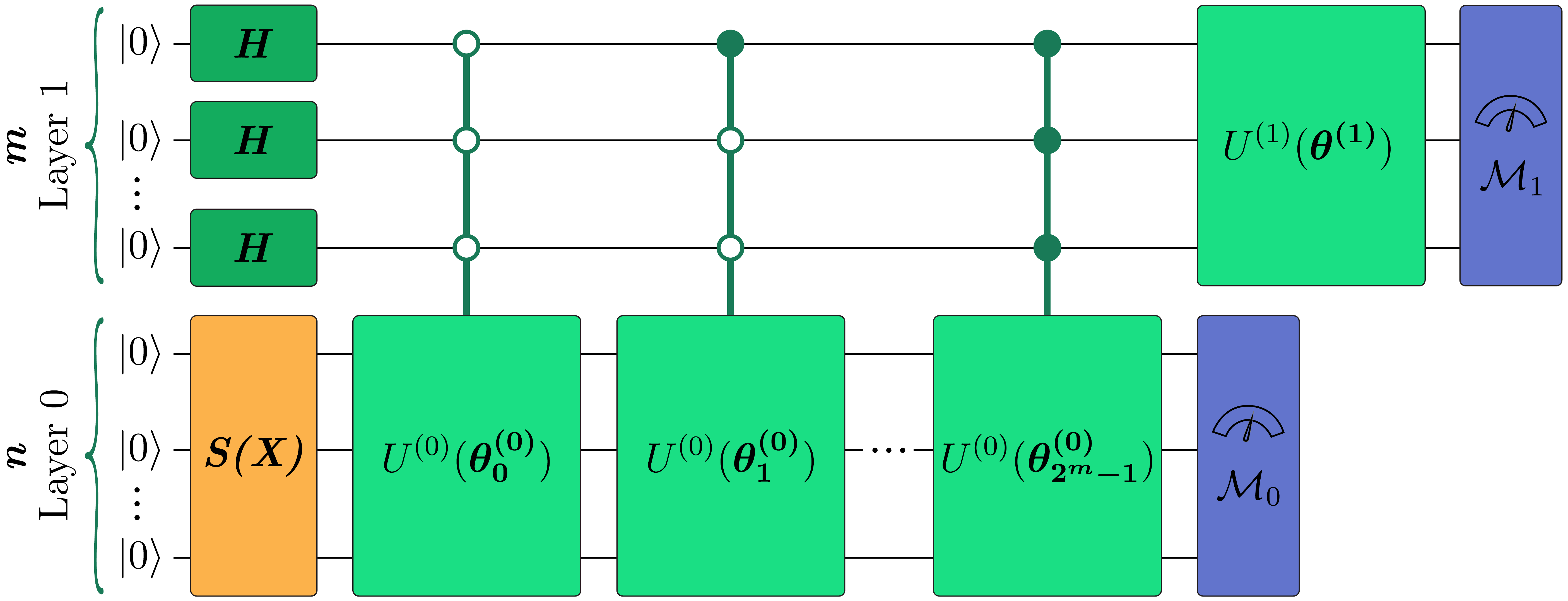}
  \caption{A SPQC circuit with $n$ qubits in the first layer and $m$ qubits in the second layer. 
           FFQRAM loads $L=2^{m}$ parameterised sub-circuits onto one 
           $n$-qubit data register, using only $m$ address qubits. 
           Black (white) controls indicate conditioning on the 
           $\ket{1}$ ($\ket{0}$) state.}
  \label{fig:ffqram-bagging}
\end{figure}

Figure~\ref{fig:ffqram-bagging} illustrates this baseline architecture: 
$L$ parameterised sub-circuits are executed in superposition on a single 
data register, but no activation layer is applied yet.

\subsection*{Introducing Non-Linearity}
RUS gearboxes supply the activation functions in an SPQC. The idea is to prepare several coherent copies of a layer, feed each copy the same input, and post-select only those shots in which every copy collapses to an identical computational-basis string. Because the projector is a binary-string observable, the FFQRAM address register remains coherent throughout the procedure. 

\begin{figure}[htbp]
  \centering
  \includegraphics[width=\textwidth]{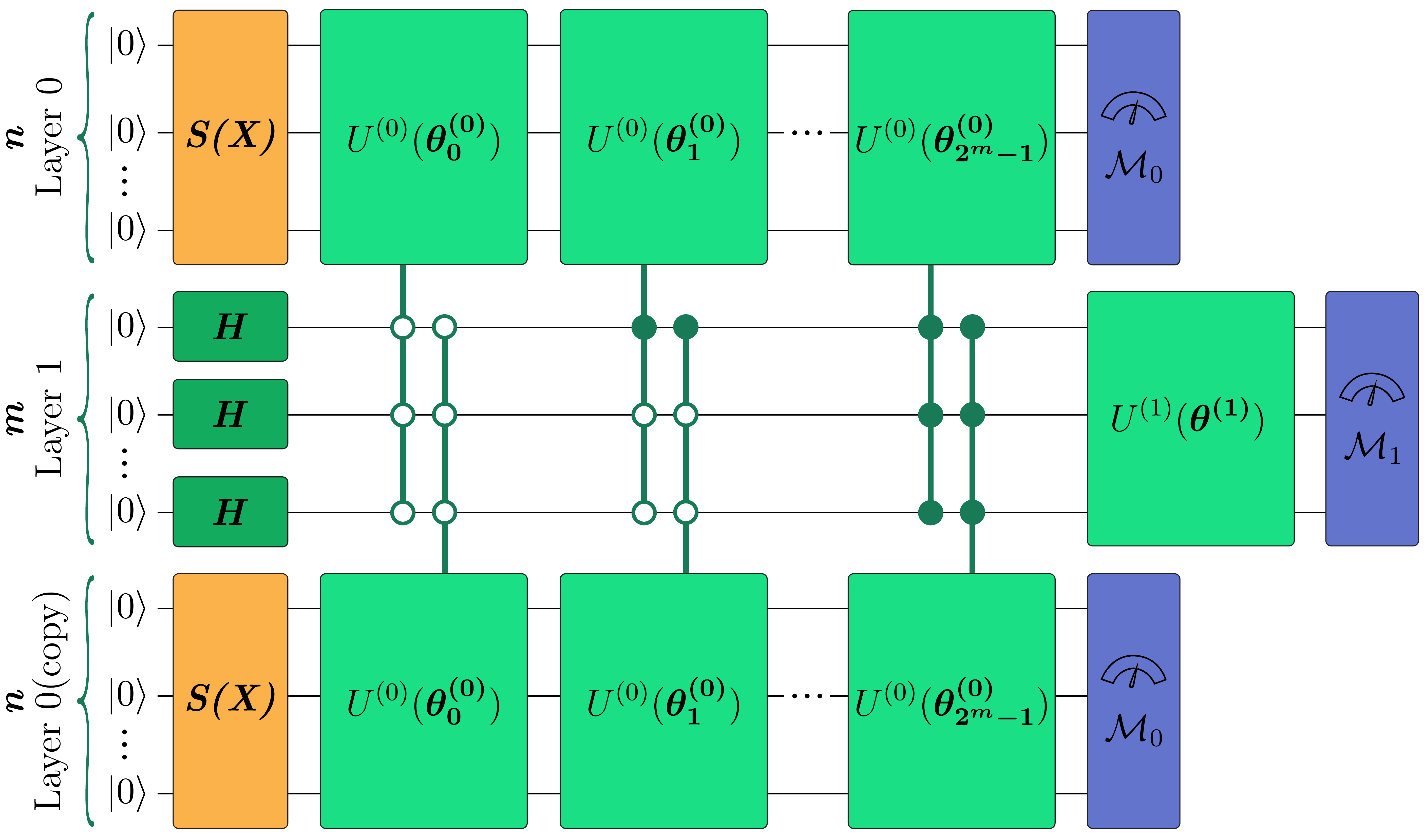}
  \caption{Two-layer SPQC with quadratic activation.  FFQRAM loads the
           parameter sets $\boldsymbol{\theta}^{(j)}$, and an RUS gearbox
           squares the first-layer outputs before they feed the second layer.
           Black (white) controls indicate conditioning on the $\ket{1}$
           ($\ket{0}$) state.}
  \label{fig:SPQC-circuit-quad}
\end{figure}

Figure~\ref{fig:SPQC-circuit-quad} shows a generalised SPQC circuit with two hidden layers, first with $2^n$ neurons and second with $2^m$. A quadratic activation function is applied to the neurons at the end of the first layer. Sec.~\ref{sec:experiments} shows the performance of this model on an artificial dataset. 

This construction can be derived in a few clear steps. First, consider a single-layer SPQC (or, equivalently, a standard PQC) acting on $n$ qubits. The circuit applies the data-encoding unitary $S(\boldsymbol{X})$, then the variational block $U(\boldsymbol{\theta})$, and finally measures an observable $\mathcal{M}$. Its expectation value is
\[
\langle \mathcal{M} \rangle
  = \bra{0}^{\otimes n}
      S^\dagger(\boldsymbol{X})
      U^\dagger(\boldsymbol{\theta})
      \mathcal{M}
      U(\boldsymbol{\theta})
      S(\boldsymbol{X})
    \ket{0}^{\otimes n}.
\]

To extend this circuit into an SPQC layer capable of parallel processing, we attach $m$ address qubits and use FFQRAM. First, Hadamard gates prepare the address register in the superposition $\frac{1}{\sqrt{L}}\sum_{j=0}^{L-1}\ket{j}$, while the original $n$ qubits receive the data unitary $S(\boldsymbol{X})$. Second, FFQRAM writes the conditional unitary $U(\boldsymbol{\theta}_j)$ onto the data register whenever the address state is $\ket{j}$; this is implemented with a cascade of multi-controlled gates \cite{Park2019Circuit}. Third, we measure the data qubits with a single-string projector, for example $\mathcal{M} = (|0\rangle\!\langle 0|)^{\otimes n}$, and retain only those shots that return the all-zero outcome. For every accepted shot the amplitude of each address state $\ket{j}$ equals the output $p_j$ of the first layer parameterised by $\boldsymbol{\theta}_j$:
\begin{align*}
  \ket{\psi}_{\text{first}}
    &= \bigl(S(\boldsymbol{X})\ket{0}^{\otimes n}\bigr)\!\otimes\!
       \bigl(H\ket{0}\bigr)^{\!\otimes m}
     = 2^{-m/2}
       \Bigl(\sum_{i=0}^{2^{n}-1}\alpha_i(\boldsymbol{X})\ket{i}\Bigr)
       \otimes
       \Bigl(\sum_{j=0}^{2^{m}-1}\ket{j}\Bigr), \\
  \ket{\psi}_{\text{second}}
    &= 2^{-m/2}\!
       \sum_{j=0}^{2^{m}-1}
       \bigl(U(\boldsymbol{\theta}_j)S(\boldsymbol{X})\ket{0}^{\otimes n}\bigr)
       \otimes\ket{j}
     = 2^{-m/2}
       \sum_{j=0}^{2^{m}-1}\sum_{i=0}^{2^{n}-1}
       \bigl(U(\boldsymbol{\theta}_j)\alpha_i(\boldsymbol{X})\ket{i}\bigr)\ket{j}, \\
  \ket{\psi}_{\text{third}}
    &= 2^{-m/2}
       \sum_{j=0}^{2^{m}-1}
         \bra{0}^{\otimes n}
           U(\boldsymbol{\theta}_j)S(\boldsymbol{X})
         \ket{0}^{\otimes n}
         \,\ket{j}
     = \frac{2^{-m/2}}{\mathcal{N}}
       \sum_{j=0}^{2^{m}-1}p_j\ket{j},
\end{align*}
where
$\displaystyle
\mathcal{N}=
\sum_{k=0}^{2^{n}-1}\sum_{j=0}^{2^{m}-1}
  \bra{k}U(\boldsymbol{\theta}_j)S(\boldsymbol{X})\ket{0}^{\otimes n}
$
normalises the post-selected state.

To introduce a non-linear activation, we duplicate the whole layer and apply the same post-selection to both copies. With two replicas the amplitudes are squared, yielding a quadratic activation:
\begin{align*}
  \ket{\psi}^{\text{quadratic}}_{\text{second}}
    &= 2^{-m/2}
       \sum_{j=0}^{2^{m}-1}\sum_{i,i'=0}^{2^{n}-1}
         \bigl(U(\boldsymbol{\theta}_j)\alpha_i(\boldsymbol{X})\ket{i}\bigr)
         \bigl(U(\boldsymbol{\theta}_j)\alpha_{i'}(\boldsymbol{X})\ket{i'}\bigr)
         \ket{j}, \\
  \ket{\psi}^{\text{quadratic}}_{\text{third}}
    &= 2^{-m/2}
       \sum_{j=0}^{2^{m}-1}
       \Bigl(
         \sum_{i=0}^{2^{n}-1}
           \bra{0}^{\otimes n}
             U(\boldsymbol{\theta}_j)\alpha_i(\boldsymbol{X})\ket{i}
       \Bigr)^{\!2}\ket{j}
     = \frac{2^{-m/2}}{\mathcal{N}^{2}}
       \sum_{j=0}^{2^{m}-1}p_j^{2}\ket{j}.
\end{align*}

The statevector amplitudes are now transformed by a quadratic function. Repeating the same procedure with additional copies implements higher-degree activations and builds deeper SPQC layers.

\section{Numerical Experiments and Results}\label{sec:experiments}

In this section, we demonstrate the effectiveness of SPQCs on one-dimensional (1D) regression and two-dimensional (2D) classification tasks. Our experiments are designed to highlight the impact of different activation functions (linear vs.\ quadratic), as well as the role of increasing the number of ancillary qubits in the FFQRAM. All models were trained using a gradient-based hybrid quantum-classical optimization scheme, and we report average results over multiple random initializations.

\subsection{One-Dimensional Regression Task}\label{subsec:1D-toy}

We begin with a synthetic step-function regression problem comprising $200$ equally spaced inputs whose target is a square wave. All models are trained with the Adam optimiser (learning rate $10^{-2}$) for 5000 epochs, and every experiment is repeated over five random seeds; reported values are means with one–standard-deviation.

\begin{figure}[ht]
  \centering
  \includegraphics[width=0.7\textwidth]{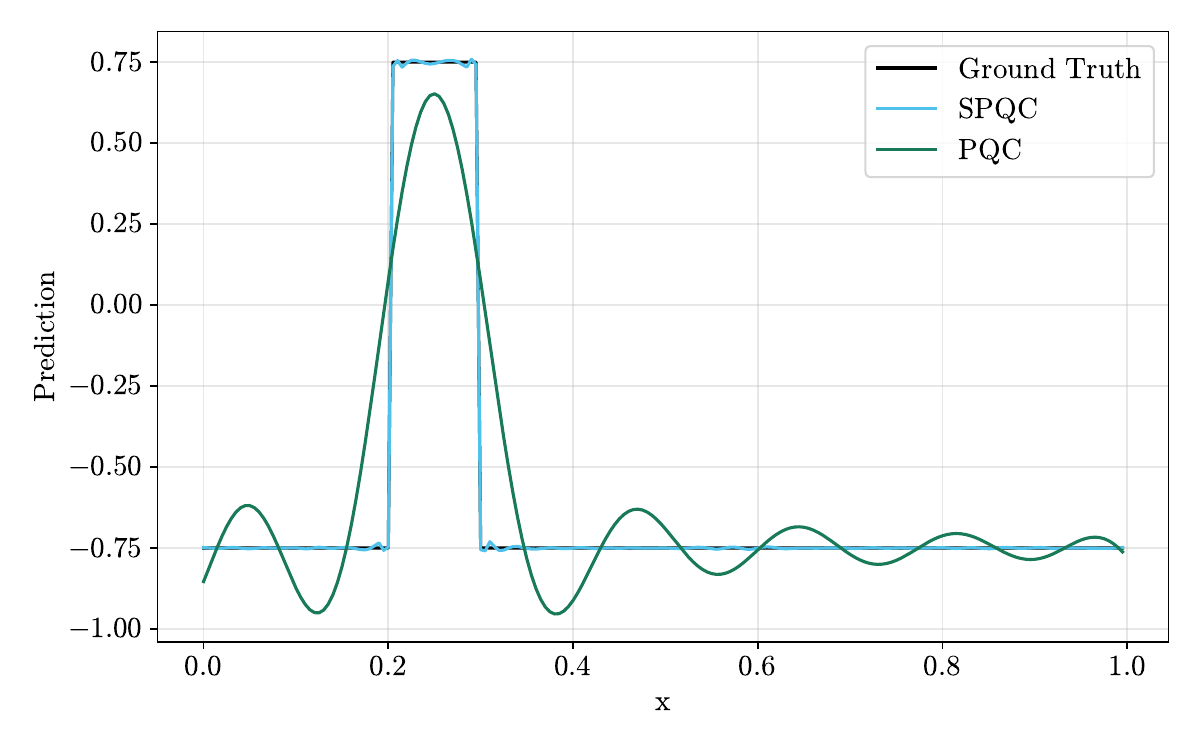}
  \caption{Step-function regression.  Black: ground truth; blue: SPQC with $n = 2, m=2$; green: depth-matched PQC.  Both models have the same total parameter count, yet the SPQC tracks the discontinuities far more accurately.}
  \label{fig:SPQC_PQC}
\end{figure}

To evaluate the practical value of parameter superposition, an SPQC with $n = 2$ data qubits and $m = 2$ flip–flop ancillas is compared against a depth-matched PQC. The ancillas index $L=4$ independent sub-models executed coherently in parallel, whereas the baseline PQC uses the same two data qubits but quadruples its circuit depth so that the total number of trainable parameters equals the combined parameter count of the four SPQC branches. Table~\ref{tab:stepfunc-metrics} and Fig.~\ref{fig:SPQC_PQC} show that, even without an explicit activation layer, the SPQC achieves an $R^{2}$ score of $1.000\pm5.1\times10^{-5}$ and an MSE three orders of magnitude lower than the conventional circuit. Parameter superposition therefore yields a markedly sharper approximation of the discontinuous target.

\begin{table}[ht]
  \centering
  \caption{Regression accuracy on the step-function task for $n = 2, m = 2$.  
The SPQC uses two flip–flop ancillas, while the PQC has the same data-qubit layout and four-times the depth to match the total parameter count.}
  \label{tab:stepfunc-metrics}
  \begin{tabular}{lccc}
    \toprule
    Model & MSE & MAE & $R^{2}$ \\
    \midrule
    SPQC           
         & $2.8\!\times\!10^{-5}\;\pm\;9.8\!\times\!10^{-6}$
         & $3.54\!\times\!10^{-3}\;\pm\;6.97\!\times\!10^{-4}$
         & $1.000\;\pm\;5.1\!\times\!10^{-5}$ \\[2pt]
    PQC                
         & $3.80\!\times\!10^{-2}\;\pm\;8.0\!\times\!10^{-5}$
         & $1.16\!\times\!10^{-1}\;\pm\;2.9\!\times\!10^{-4}$
         & $8.04\!\times\!10^{-1}\;\pm\;4.1\!\times\!10^{-4}$ \\
    \bottomrule
  \end{tabular}
\end{table}

\begin{figure}[ht]
  \centering
  \includegraphics[width=1.0\textwidth]{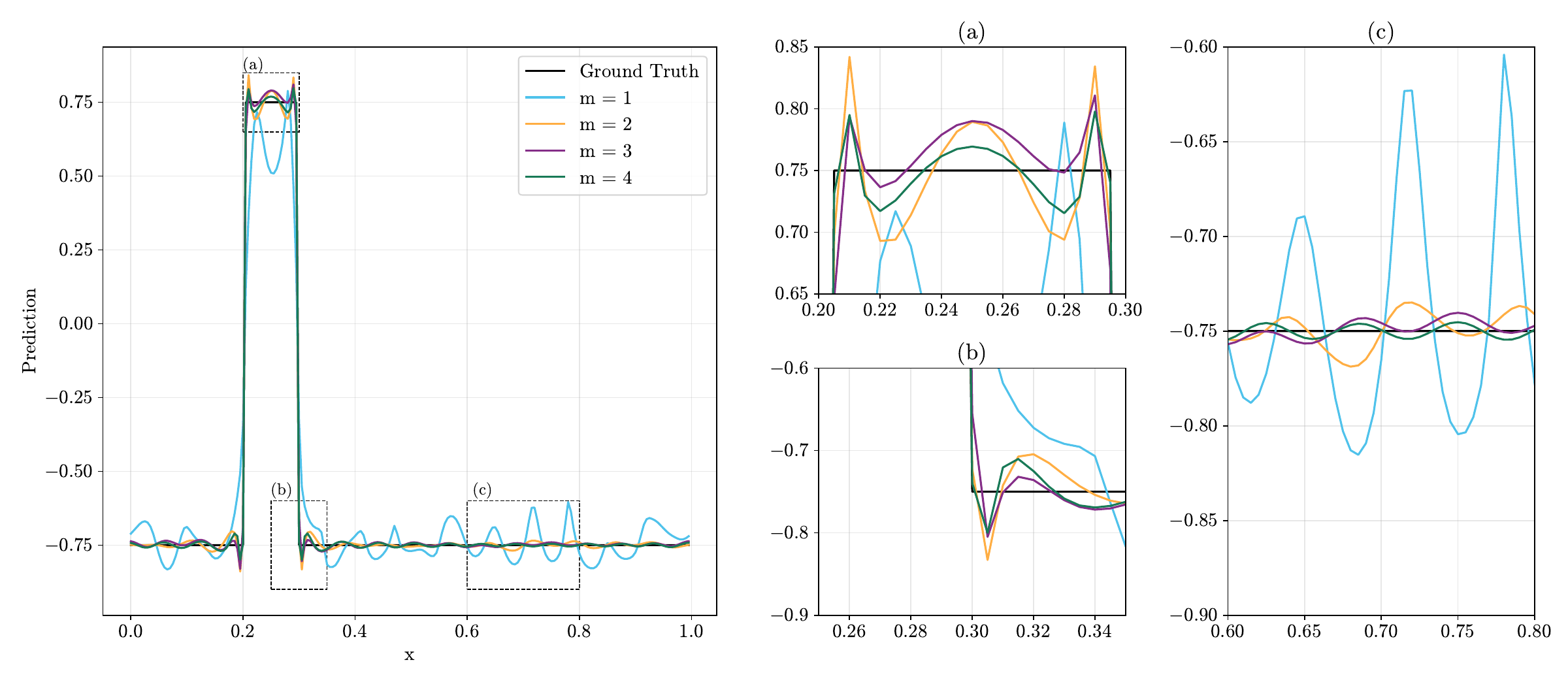}
  \caption{Mean SPQC predictions with a linear activation and $m=1,2,3,4$ flip–flop ancillas (five seeds each).  Insets zoom in on the discontinuities; the sharpest transitions arise for $m=4$.}
  \label{fig:linear_case_num_flip_flops}
\end{figure}

Having established a baseline advantage, we examine how expressivity scales with the number of flip–flop ancillas. For a single data qubit ($n=1$) we vary the address-register size $m=1,2,3,4$ while retaining a linear activation. Each additional ancilla doubles the size of the coherent ensemble. Figure~\ref{fig:linear_case_num_flip_flops} depicts the mean predictions, and Table~\ref{tab:m-scan} lists quantitative metrics. The error drops precipitously when moving from $m=1$ to $m=2$, showing that a logarithmic increase in qubits can dramatically enhance performance. Improvements beyond $m=2$ are non-monotonic, yet the best overall accuracy is reached at $m=4$, where both bias and variance are lowest.

\begin{table}[ht]
  \centering
  \caption{Effect of the number of flip–flop ancillas $m$ on step-function regression accuracy for a model with $n=1$.}
  \label{tab:m-scan}
  \begin{tabular}{ccc}
    \toprule
    $m$ & MSE & MAE \\
    \midrule
    1 & $2.77\times10^{-2}\ \pm\ 3.63\times10^{-2}$ & $9.33\times10^{-2}\ \pm\ 7.15\times10^{-2}$ \\
    2 & $5.89\times10^{-4}\ \pm\ 4.06\times10^{-4}$ & $1.57\times10^{-2}\ \pm\ 6.18\times10^{-3}$ \\
    3 & $1.08\times10^{-3}\ \pm\ 1.91\times10^{-3}$ & $1.39\times10^{-2}\ \pm\ 1.02\times10^{-2}$ \\
    4 & $1.63\times10^{-4}\ \pm\ 8.01\times10^{-5}$ & $7.95\times10^{-3}\ \pm\ 1.81\times10^{-3}$ \\
    \bottomrule
  \end{tabular}
\end{table}

\subsection{Two-Dimensional Classification Task}\label{subsec:2D-task}

We next consider a two-dimensional star-shaped classification problem whose decision boundary is highly non-linear \cite{Bishop1995Neural,Goodfellow2016Deep}. Each point $(x,y)$ is labelled according to whether it lies inside or outside a five-arm star polygon, a geometry that poses a challenge for models restricted to linear amplitude maps. Two SPQC variants are compared: one with a linear activation and one with a quadratic activation obtained by duplicating the functional register and post-selecting on equal outcomes.

\begin{figure}[ht]
  \centering
  \includegraphics[width=\textwidth]{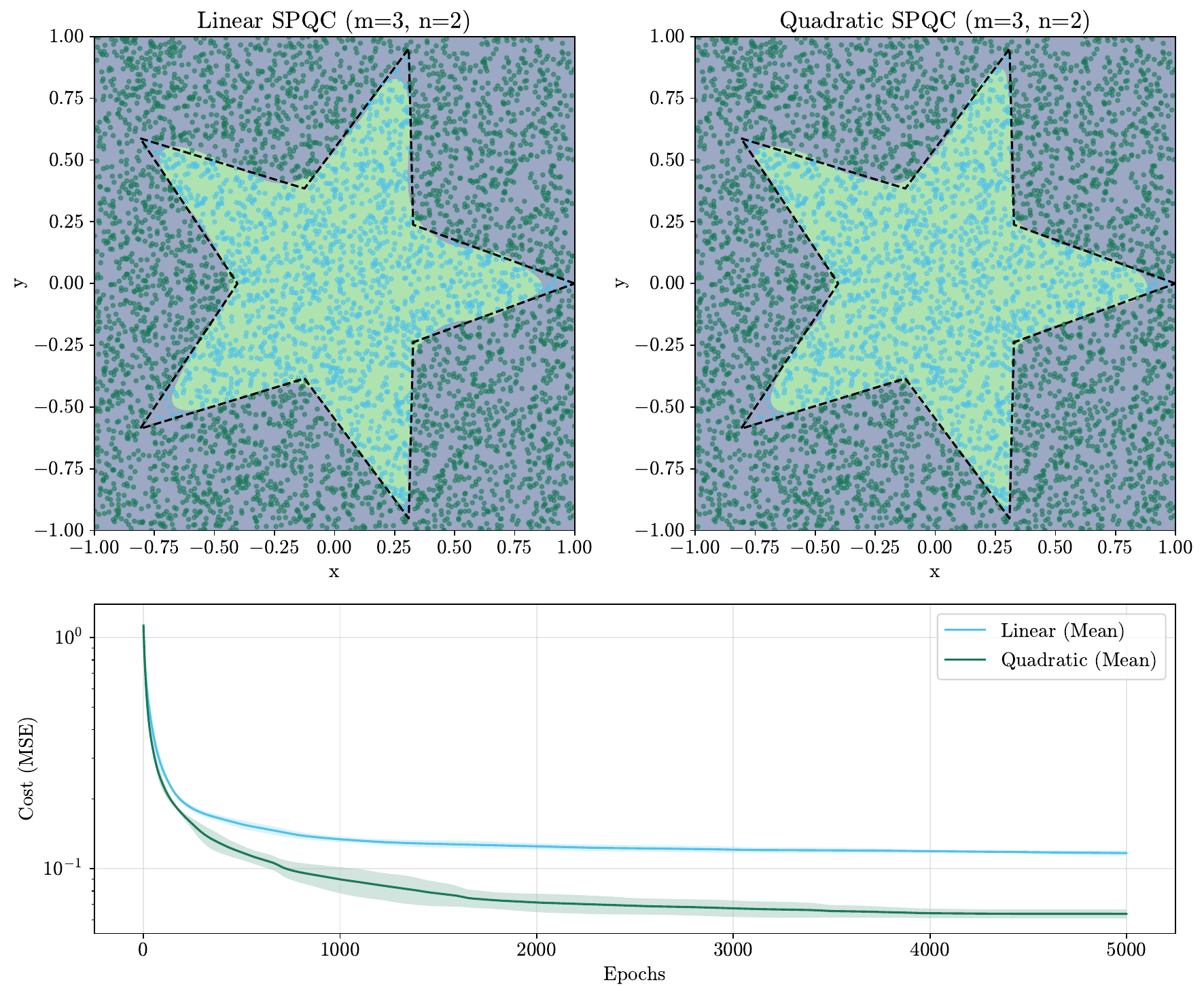}
  \caption{Star-shaped classification.  Top: decision boundaries learned by the linear (left) and quadratic (right) SPQCs; the black dashed curve is the true boundary.  Bottom: training-loss curves averaged over five seeds.  The quadratic activation aligns more closely with the complex geometry and converges to a lower final loss.}
  \label{fig:linear_quad_SPQC}
\end{figure}

Both models use three flip–flop address qubits ($m=3$), giving $L=8$ coherent sub-models. The linear version employs two data qubits, for a total of five qubits; the quadratic version adds two more data qubits to host the duplicate register, for a total of seven. Each network is trained for the same number of epochs with the Adam optimiser (learning rate $10^{-2}$) and evaluated over five random seeds. Table~\ref{tab:star-metrics} summarises the results, and Fig.~\ref{fig:linear_quad_SPQC} shows the learned decision boundaries alongside the training-loss trajectories.

The quadratic SPQC reduces the mean-absolute error by roughly eight percent (0.436 → 0.401) and boosts classification accuracy by about one percentage point, while simultaneously cutting the accuracy standard deviation by a factor of three, indicating more stable convergence. Its mean-squared error rises slightly, but the $\sim1.5\,\%$ difference is well within one standard deviation and therefore not statistically significant. These findings suggest that introducing a quadratic non-linearity improves fidelity near the intricate star edges without incurring a substantial qubit overhead.

\begin{table}[ht]
  \centering
  \caption{Star-dataset classification with three address qubits ($m=3$), averaged over five seeds.  The quadratic model duplicates the functional register, adding a quadratic amplitude transformation.  Metrics are reported as mean $\pm$ standard deviation.}
  \label{tab:star-metrics}
  \begin{tabular}{lccc}
    \toprule
    Model & MSE & MAE & Accuracy [\%] \\
    \midrule
    Linear SPQC      & $0.647 \,\pm\,0.007$ & $0.436 \,\pm\,0.005$ & $80.37 \,\pm\,0.15$ \\
    Quadratic SPQC   & $0.657 \,\pm\,0.013$ & $0.401 \,\pm\,0.004$ & $81.38 \,\pm\,0.06$ \\
    \bottomrule
  \end{tabular}
\end{table}

\section{Discussion}\label{sec:discussion}

SPQC address two core obstacles in variational QML. By loading \(L=2^{m}\) independent parameter sets into superposition with only \(m\) flip–flop ancillas, they avoid the linear qubit blow-up that would otherwise accompany an ensemble. In addition, RUS post-selection supplies polynomial activations, allowing SPQCs to move beyond the strictly linear decision boundaries that confine ordinary PQCs. The numerical results make both advantages visible: superposition alone outperforms a depth-matched PQC on a discontinuous regression task, and introducing a quadratic activation yields higher accuracy and lower variance on a challenging two-dimensional classification problem.

From a hardware-resource perspective the circuit depth and qubit count remain modest. With RUS-optimised synthesis the architecture requires only \(n+\log L\) qubits and achieves a non-Clifford depth that scales as \(O(\log n)\) \cite{Duan2024Compact,Guerreschi2019RUS}. Coherently evaluating all \(L\) sub-models therefore trades the classical overhead of repeated circuit initialisation for a small ancilla register, a sensible exchange on NISQ hardware where reset times dominate wall-clock runtime. Simulations furthermore indicate that circuit-based QRAM outperforms bucket-brigade designs under realistic noise assumptions \cite{Connor2021ResilienceQRAM}, suggesting that the memory primitive itself is within near-term reach.

The primary limitation is the shot overhead introduced by post-selection. If \(p_{\mathrm{succ}}\) denotes the single-copy success probability and \(r\) copies realise an \(r\)-th-order activation, the retention rate scales as \(\alpha\approx p_{\mathrm{succ}}^{\,r}\) \cite{Paetznick2013RUS}. Higher-degree activations thus demand either more shots or amplitude-amplification routines, while each additional copy deepens the circuit through multi-controlled gates, increasing vulnerability to decoherence.

Several techniques could mitigate these costs. Fixed-point amplitude amplification can boost \(p_{\mathrm{succ}}\) without large depth penalties, and lightweight error-correction or error-mitigation could extend coherent runtime. Partial post-selection schemes that recycle near-miss outcomes, together with noise-aware compilations that compress the FFQRAM cascade, may further reduce the effective sample complexity. Determining the best combination of these tools for specific hardware platforms is a natural next step.

By combining ensemble-style parameter superposition with trainable non-linearities, SPQCs bring two pillars of classical deep learning—bagging and activation functions—into a single quantum circuit. The result is a template for deeper, more expressive quantum neural networks that does not impose an exponential qubit cost. Future work should test the architecture on larger, real-world data sets, experiment with higher-order and learned activations, and refine compilation and sampling techniques so that practical performance keeps pace with the theoretical gains outlined here.

\section{Conclusion and Future Works}\label{sec:conclusion}

We have presented SPQCs as a novel framework that integrates FFQRAM-based parallelization with RUS activations. This synergy extends beyond standard PQCs by simultaneously addressing multi-output scalability and the lack of explicit non-linearities—two major limitations in QML. Numerical evaluations on both one-dimensional regression and two-dimensional classification problems indicate that SPQCs can outperform linear PQCs, especially when additional ancillary qubits or higher-order polynomial layers are employed.

Future work can expand in several directions. First, refining RUS protocols to reduce shot overhead and post-selection failures could ease the demands on near-term quantum devices. Second, investigating partial post-selection strategies or approximate polynomial layers may further mitigate resource costs. Third, the inclusion of error mitigation or quantum error correction methods aligned with FFQRAM-based routing could bolster circuit fidelity against hardware noise \cite{Cerezo2021VQA,kuzmin2025method}. Finally, adapting SPQCs to more advanced data-encoding schemes or larger problem instances — where true quantum advantage may emerge — remains an important milestone. We anticipate that continued research on multi-layered, non-linear quantum architectures will play a central role in realizing robust and expressive quantum models in practical machine learning applications.

\bibliography{sn-bibliography}

\begin{thebibliography}{10}
\expandafter\ifx\csname url\endcsname\relax
  \def\url#1{\burl{#1}}\fi
\expandafter\ifx\csname urlprefix\endcsname\relax\def\urlprefix{URL }\fi
\providecommand{\bibinfo}[2]{#2}
\providecommand{\eprint}[2][]{\url{#2}}
\providecommand{\doi}[1]{\url{https://doi.org/#1}}
\bibcommenthead

\bibitem{Peruzzo2014Variational}
\bibinfo{author}{Peruzzo, A.} \emph{et~al.}
\newblock \bibinfo{title}{A variational eigenvalue solver on a quantum processor}.
\newblock \emph{\bibinfo{journal}{Nature Communications}} \textbf{\bibinfo{volume}{5}}, \bibinfo{pages}{4213} (\bibinfo{year}{2014}).

\bibitem{Tilly2022Variational}
\bibinfo{author}{Tilly, J.} \emph{et~al.}
\newblock \bibinfo{title}{The variational quantum eigensolver: A review of methods and best practices}.
\newblock \emph{\bibinfo{journal}{Physics Reports}} \textbf{\bibinfo{volume}{986}}, \bibinfo{pages}{1–128} (\bibinfo{year}{2022}).

\bibitem{Melnikov2023Quantum}
\bibinfo{author}{Melnikov, A.}, \bibinfo{author}{Kordzanganeh, M.}, \bibinfo{author}{Alodjants, A.} \& \bibinfo{author}{Lee, R.-K.}
\newblock \bibinfo{title}{{Quantum machine learning: from physics to software engineering}}.
\newblock \emph{\bibinfo{journal}{Advances in Physics: X}} \textbf{\bibinfo{volume}{8}}, \bibinfo{pages}{2165452} (\bibinfo{year}{2023}).

\bibitem{Benedetti2019Parameterized}
\bibinfo{author}{Benedetti, M.}, \bibinfo{author}{Lloyd, E.}, \bibinfo{author}{Sack, S.} \& \bibinfo{author}{Fiorentini, M.}
\newblock \bibinfo{title}{{Parameterized quantum circuits as machine learning models}}.
\newblock \emph{\bibinfo{journal}{Quantum Science and Technology}} \textbf{\bibinfo{volume}{4}}, \bibinfo{pages}{043001} (\bibinfo{year}{2019}).

\bibitem{Schuld2019Quantum}
\bibinfo{author}{Schuld, M.} \& \bibinfo{author}{Killoran, N.}
\newblock \bibinfo{title}{{Quantum Machine Learning in Feature Hilbert Spaces}}.
\newblock \emph{\bibinfo{journal}{Physical Review Letters}} \textbf{\bibinfo{volume}{122}}, \bibinfo{pages}{040504} (\bibinfo{year}{2019}).

\bibitem{Schuld2021Effect}
\bibinfo{author}{Schuld, M.}, \bibinfo{author}{Sweke, R.} \& \bibinfo{author}{Meyer, J.~J.}
\newblock \bibinfo{title}{{Effect of data encoding on the expressive power of variational quantum-machine-learning models}}.
\newblock \emph{\bibinfo{journal}{Physical Review A}} \textbf{\bibinfo{volume}{103}}, \bibinfo{pages}{032430} (\bibinfo{year}{2021}).

\bibitem{Abbas2021Power}
\bibinfo{author}{Abbas, A.} \emph{et~al.}
\newblock \bibinfo{title}{The power of quantum neural networks}.
\newblock \emph{\bibinfo{journal}{Nature Computational Science}} \textbf{\bibinfo{volume}{1}}, \bibinfo{pages}{403--409} (\bibinfo{year}{2021}).

\bibitem{Caro2022Generalization}
\bibinfo{author}{Caro, M.~C.} \emph{et~al.}
\newblock \bibinfo{title}{{Generalization in quantum machine learning from few training data}}.
\newblock \emph{\bibinfo{journal}{Nature Communications}} \textbf{\bibinfo{volume}{13}}, \bibinfo{pages}{4919} (\bibinfo{year}{2022}).

\bibitem{Kordzanganeh2021Quantum}
\bibinfo{author}{Kordzanganeh, M.}, \bibinfo{author}{Utting, A.} \& \bibinfo{author}{Scaife, A.}
\newblock \bibinfo{title}{{Quantum Machine Learning for Radio Astronomy}}.
\newblock \emph{\bibinfo{journal}{arXiv preprint arXiv: 2112.026555}}  (\bibinfo{year}{2021}).

\bibitem{sagingalieva2025hybrid}
\bibinfo{author}{Sagingalieva, A.}, \bibinfo{author}{Lusnig, L.}, \bibinfo{author}{Cavalli, F.} \& \bibinfo{author}{Melnikov, A.}
\newblock \bibinfo{title}{Hybrid quantum neural networks for computer-aided sex diagnosis in forensic and physical anthropology}.
\newblock \emph{\bibinfo{journal}{Informatics in Medicine Unlocked}} \textbf{\bibinfo{volume}{58}}, \bibinfo{pages}{101682} (\bibinfo{year}{2025}).

\bibitem{kurkin2025forecasting}
\bibinfo{author}{Kurkin, A.}, \bibinfo{author}{Hegemann, J.}, \bibinfo{author}{Kordzanganeh, M.} \& \bibinfo{author}{Melnikov, A.}
\newblock \bibinfo{title}{Forecasting steam mass flow in power plants using the parallel hybrid network}.
\newblock \emph{\bibinfo{journal}{Engineering Applications of Artificial Intelligence}} \textbf{\bibinfo{volume}{160}}, \bibinfo{pages}{111912} (\bibinfo{year}{2025}).

\bibitem{Havlicek2019Supervised}
\bibinfo{author}{Havlíček, V.} \emph{et~al.}
\newblock \bibinfo{title}{{Supervised learning with quantum-enhanced feature spaces}}.
\newblock \emph{\bibinfo{journal}{Nature}} \textbf{\bibinfo{volume}{567}}, \bibinfo{pages}{209–212} (\bibinfo{year}{2019}).

\bibitem{Schuld2018Supervised}
\bibinfo{author}{Schuld, M.} \& \bibinfo{author}{Petruccione, F.}
\newblock \emph{\bibinfo{title}{Supervised Learning with Quantum Computers}} Quantum Science and Technology (\bibinfo{publisher}{Springer, Cham}, \bibinfo{year}{2018}).

\bibitem{Huang2021Power}
\bibinfo{author}{Huang, H.-Y.} \emph{et~al.}
\newblock \bibinfo{title}{{Power of data in quantum machine learning}}.
\newblock \emph{\bibinfo{journal}{Nature Communications}} \textbf{\bibinfo{volume}{12}}, \bibinfo{pages}{2631} (\bibinfo{year}{2021}).

\bibitem{Goodfellow2016Deep}
\bibinfo{author}{Goodfellow, I.}, \bibinfo{author}{Bengio, Y.} \& \bibinfo{author}{Courville, A.}
\newblock \emph{\bibinfo{title}{Deep Learning}}  (\bibinfo{publisher}{MIT Press}, \bibinfo{year}{2016}).

\bibitem{McClean2018Barren}
\bibinfo{author}{McClean, J.~R.}, \bibinfo{author}{Boixo, S.}, \bibinfo{author}{Smelyanskiy, V.~N.}, \bibinfo{author}{Babbush, R.} \& \bibinfo{author}{Neven, H.}
\newblock \bibinfo{title}{Barren plateaus in quantum neural network training landscapes}.
\newblock \emph{\bibinfo{journal}{Nature Communications}} \textbf{\bibinfo{volume}{9}}, \bibinfo{pages}{4812} (\bibinfo{year}{2018}).

\bibitem{Zhao2021Analyzing}
\bibinfo{author}{Zhao, C.} \& \bibinfo{author}{Gao, X.-S.}
\newblock \bibinfo{title}{Analyzing the barren plateau phenomenon in training quantum neural networks with the {ZX}-calculus}.
\newblock \emph{\bibinfo{journal}{{Quantum}}} \textbf{\bibinfo{volume}{5}}, \bibinfo{pages}{466} (\bibinfo{year}{2021}).

\bibitem{You2021Exponentially}
\bibinfo{author}{You, X.} \& \bibinfo{author}{Wu, X.}
\newblock \emph{\bibinfo{title}{Exponentially many local minima in quantum neural networks}}, \bibinfo{pages}{12144--12155} (\bibinfo{organization}{PMLR}, \bibinfo{year}{2021}).

\bibitem{Kordzanganeh2023Benchmarking}
\bibinfo{author}{Kordzanganeh, M.} \emph{et~al.}
\newblock \bibinfo{title}{Benchmarking simulated and physical quantum processing units using quantum and hybrid algorithms}.
\newblock \emph{\bibinfo{journal}{Advanced Quantum Technologies}} \textbf{\bibinfo{volume}{6}}, \bibinfo{pages}{2300043} (\bibinfo{year}{2023}).

\bibitem{Schuld2022Quantum}
\bibinfo{author}{Schuld, M.} \& \bibinfo{author}{Killoran, N.}
\newblock \bibinfo{title}{Is quantum advantage the right goal for quantum machine learning?}
\newblock \emph{\bibinfo{journal}{PRX Quantum}} \textbf{\bibinfo{volume}{3}}, \bibinfo{pages}{030101} (\bibinfo{year}{2022}).

\bibitem{Park2019Circuit}
\bibinfo{author}{Park, D.~K.}, \bibinfo{author}{Petruccione, F.} \& \bibinfo{author}{Rhee, J.-K.~K.}
\newblock \bibinfo{title}{Circuit-based quantum random access memory for classical data}.
\newblock \emph{\bibinfo{journal}{Scientific Reports}} \textbf{\bibinfo{volume}{9}} (\bibinfo{year}{2019}).

\bibitem{Koppe2023Amplitude}
\bibinfo{author}{Koppe, J.} \& \bibinfo{author}{Wolf, M.-O.}
\newblock \bibinfo{title}{Amplitude-based implementation of the unit step function on a quantum computer}.
\newblock \emph{\bibinfo{journal}{Physical Review A}} \textbf{\bibinfo{volume}{107}}, \bibinfo{pages}{022606} (\bibinfo{year}{2023}).

\bibitem{Krizhevsky2012ImageNet}
\bibinfo{author}{Krizhevsky, A.}, \bibinfo{author}{Sutskever, I.} \& \bibinfo{author}{Hinton, G.~E.}
\newblock \bibinfo{editor}{Pereira, F.}, \bibinfo{editor}{Burges, C.}, \bibinfo{editor}{Bottou, L.} \& \bibinfo{editor}{Weinberger, K.} (eds) \emph{\bibinfo{title}{Imagenet classification with deep convolutional neural networks}}.
\newblock (eds \bibinfo{editor}{Pereira, F.}, \bibinfo{editor}{Burges, C.}, \bibinfo{editor}{Bottou, L.} \& \bibinfo{editor}{Weinberger, K.}) \emph{\bibinfo{booktitle}{Advances in Neural Information Processing Systems}}, Vol.~\bibinfo{volume}{25} (\bibinfo{publisher}{Curran Associates, Inc.}, \bibinfo{year}{2012}).

\bibitem{He2015Deep}
\bibinfo{author}{He, K.}, \bibinfo{author}{Zhang, X.}, \bibinfo{author}{Ren, S.} \& \bibinfo{author}{Sun, J.}
\newblock \emph{\bibinfo{title}{Deep residual learning for image recognition}}, \bibinfo{pages}{770--778} (\bibinfo{year}{2016}).

\bibitem{Simonyan2015Very}
\bibinfo{author}{Simonyan, K.} \& \bibinfo{author}{Zisserman, A.}
\newblock \bibinfo{title}{Very deep convolutional networks for large-scale image recognition}.
\newblock \emph{\bibinfo{journal}{arXiv preprint arXiv:1409.1556}}  (\bibinfo{year}{2014}).

\bibitem{Vaswani2017Attention}
\bibinfo{author}{Vaswani, A.} \emph{et~al.}
\newblock \bibinfo{editor}{Guyon, I.} \emph{et~al.} (eds) \emph{\bibinfo{title}{Attention is all you need}}.
\newblock (eds \bibinfo{editor}{Guyon, I.} \emph{et~al.}) \emph{\bibinfo{booktitle}{Advances in Neural Information Processing Systems}}, Vol.~\bibinfo{volume}{30} (\bibinfo{publisher}{Curran Associates, Inc.}, \bibinfo{year}{2017}).

\bibitem{Devlin2019Bert}
\bibinfo{author}{Devlin, J.}, \bibinfo{author}{Chang, M.-W.}, \bibinfo{author}{Lee, K.} \& \bibinfo{author}{Toutanova, K.}
\newblock \bibinfo{editor}{Burstein, J.}, \bibinfo{editor}{Doran, C.} \& \bibinfo{editor}{Solorio, T.} (eds) \emph{\bibinfo{title}{{BERT}: Pre-training of deep bidirectional transformers for language understanding}}.
\newblock (eds \bibinfo{editor}{Burstein, J.}, \bibinfo{editor}{Doran, C.} \& \bibinfo{editor}{Solorio, T.}) \emph{\bibinfo{booktitle}{Proceedings of the 2019 Conference of the North {A}merican Chapter of the Association for Computational Linguistics: Human Language Technologies, Volume 1 (Long and Short Papers)}}, \bibinfo{pages}{4171--4186} (\bibinfo{publisher}{Association for Computational Linguistics}, \bibinfo{address}{Minneapolis, Minnesota}, \bibinfo{year}{2019}).

\bibitem{Brown2020Language}
\bibinfo{author}{Brown, T.} \emph{et~al.}
\newblock \bibinfo{editor}{Larochelle, H.}, \bibinfo{editor}{Ranzato, M.}, \bibinfo{editor}{Hadsell, R.}, \bibinfo{editor}{Balcan, M.} \& \bibinfo{editor}{Lin, H.} (eds) \emph{\bibinfo{title}{Language models are few-shot learners}}.
\newblock (eds \bibinfo{editor}{Larochelle, H.}, \bibinfo{editor}{Ranzato, M.}, \bibinfo{editor}{Hadsell, R.}, \bibinfo{editor}{Balcan, M.} \& \bibinfo{editor}{Lin, H.}) \emph{\bibinfo{booktitle}{Advances in Neural Information Processing Systems}}, Vol.~\bibinfo{volume}{33}, \bibinfo{pages}{1877--1901} (\bibinfo{publisher}{Curran Associates, Inc.}, \bibinfo{year}{2020}).

\bibitem{Oord2016Wavenet}
\bibinfo{author}{Van Den~Oord, A.} \emph{et~al.}
\newblock \bibinfo{title}{Wavenet: A generative model for raw audio}.
\newblock \emph{\bibinfo{journal}{arXiv preprint arXiv:1609.03499}} \textbf{\bibinfo{volume}{12}} (\bibinfo{year}{2016}).

\bibitem{Shen2018Natural}
\bibinfo{author}{Shen, J.} \emph{et~al.}
\newblock \emph{\bibinfo{title}{Natural {TTS} synthesis by conditioning wavenet on {MEL} spectrogram predictions}}, \bibinfo{pages}{4779--4783} (\bibinfo{publisher}{IEEE}, \bibinfo{address}{Calgary, Canada}, \bibinfo{year}{2018}).

\bibitem{rosenblatt1958perceptron}
\bibinfo{author}{Rosenblatt, F.}
\newblock \bibinfo{title}{{The perceptron: A probabilistic model for information storage and organization in the brain.}}
\newblock \emph{\bibinfo{journal}{Psychological Review}} \textbf{\bibinfo{volume}{65}}, \bibinfo{pages}{386--408} (\bibinfo{year}{1958}).

\bibitem{werbos1974beyond}
\bibinfo{author}{Werbos, P.~J.}
\newblock \emph{\bibinfo{title}{Beyond Regression: New Tools for Prediction and Analysis in the Behavioral Sciences}}.
\newblock Ph.D. thesis, \bibinfo{school}{Harvard University} (\bibinfo{year}{1974}).

\bibitem{rumelhart1986learning}
\bibinfo{author}{Rumelhart, D.~E.}, \bibinfo{author}{Hinton, G.~E.} \& \bibinfo{author}{Williams, R.~J.}
\newblock \bibinfo{title}{Learning representations by back-propagating errors}.
\newblock \emph{\bibinfo{journal}{Nature}} \textbf{\bibinfo{volume}{323}}, \bibinfo{pages}{533--536} (\bibinfo{year}{1986}).

\bibitem{Hornik1989Multilayer}
\bibinfo{author}{Hornik, K.}, \bibinfo{author}{Stinchcombe, M.} \& \bibinfo{author}{White, H.}
\newblock \bibinfo{title}{Multilayer feedforward networks are universal approximators}.
\newblock \emph{\bibinfo{journal}{Neural Networks}} \textbf{\bibinfo{volume}{2}}, \bibinfo{pages}{359–366} (\bibinfo{year}{1989}).

\bibitem{Cybenko2012Approximation}
\bibinfo{author}{Cybenko, G.}
\newblock \bibinfo{title}{{Approximation by superpositions of a sigmoidal function}}.
\newblock \emph{\bibinfo{journal}{Mathematics of Control, Signals, and Systems (MCSS)}} \textbf{\bibinfo{volume}{2}}, \bibinfo{pages}{303--314} (\bibinfo{year}{1989}).

\bibitem{Telgarsky2016Benefits}
\bibinfo{author}{Telgarsky, M.}
\newblock \bibinfo{title}{Benefits of depth in neural networks} (\bibinfo{year}{2016}).

\bibitem{Mhaskar2016DeepVsShallow}
\bibinfo{author}{Mhaskar, H.~N.} \& \bibinfo{author}{Poggio, T.}
\newblock \bibinfo{title}{Deep vs. shallow networks: An approximation theory perspective}.
\newblock \emph{\bibinfo{journal}{Analysis and Applications}} \textbf{\bibinfo{volume}{14}}, \bibinfo{pages}{829--848} (\bibinfo{year}{2016}).

\bibitem{Poggio2017TheoryDeep}
\bibinfo{author}{Poggio, T.}, \bibinfo{author}{Mhaskar, H.}, \bibinfo{author}{Rosasco, L.}, \bibinfo{author}{Miranda, B.} \& \bibinfo{author}{Liao, Q.}
\newblock \bibinfo{title}{Why and when can deep-but not shallow-networks avoid the curse of dimensionality: a review}.
\newblock \emph{\bibinfo{journal}{International Journal of Automation and Computing}} \textbf{\bibinfo{volume}{14}}, \bibinfo{pages}{503--519} (\bibinfo{year}{2017}).

\bibitem{hinton1986learnin}
\bibinfo{author}{Paccanaro, A.} \& \bibinfo{author}{Hinton, G.~E.}
\newblock \bibinfo{title}{Learning distributed representations of concepts using linear relational embedding}.
\newblock \emph{\bibinfo{journal}{IEEE Transactions on Knowledge and Data Engineering}} \textbf{\bibinfo{volume}{13}}, \bibinfo{pages}{232--244} (\bibinfo{year}{2001}).

\bibitem{bengio2013representation}
\bibinfo{author}{Bengio, Y.}, \bibinfo{author}{Courville, A.} \& \bibinfo{author}{Vincent, P.}
\newblock \bibinfo{title}{Representation learning: A review and new perspectives}.
\newblock \emph{\bibinfo{journal}{IEEE transactions on pattern analysis and machine intelligence}} \textbf{\bibinfo{volume}{35}}, \bibinfo{pages}{1798--1828} (\bibinfo{year}{2013}).

\bibitem{kingma2014adam}
\bibinfo{author}{Kingma, D.~P.}
\newblock \bibinfo{title}{Adam: A method for stochastic optimization}.
\newblock \emph{\bibinfo{journal}{arXiv preprint arXiv:1412.6980}}  (\bibinfo{year}{2014}).

\bibitem{Ioffe2015BatchNorm}
\bibinfo{author}{Ioffe, S.} \& \bibinfo{author}{Szegedy, C.}
\newblock \emph{\bibinfo{title}{Batch normalization: Accelerating deep network training by reducing internal covariate shift}}, Vol.~\bibinfo{volume}{37} of \emph{\bibinfo{series}{Proceedings of Machine Learning Research}}, \bibinfo{pages}{448--456} (\bibinfo{publisher}{PMLR}, \bibinfo{address}{Lille, France}, \bibinfo{year}{2015}).

\bibitem{srivastava2014dropout}
\bibinfo{author}{Srivastava, N.}, \bibinfo{author}{Hinton, G.}, \bibinfo{author}{Krizhevsky, A.}, \bibinfo{author}{Sutskever, I.} \& \bibinfo{author}{Salakhutdinov, R.}
\newblock \bibinfo{title}{Dropout: A simple way to prevent neural networks from overfitting}.
\newblock \emph{\bibinfo{journal}{Journal of Machine Learning Research}} \textbf{\bibinfo{volume}{15}}, \bibinfo{pages}{1929--1958} (\bibinfo{year}{2014}).

\bibitem{nair2010relu}
\bibinfo{author}{Nair, V.} \& \bibinfo{author}{Hinton, G.~E.}
\newblock \emph{\bibinfo{title}{Rectified linear units improve restricted boltzmann machines}}, ICML'10, \bibinfo{pages}{807–814} (\bibinfo{publisher}{Omnipress}, \bibinfo{address}{Madison, WI, USA}, \bibinfo{year}{2010}).

\bibitem{he2016resnet}
\bibinfo{author}{He, K.}, \bibinfo{author}{Zhang, X.}, \bibinfo{author}{Ren, S.} \& \bibinfo{author}{Sun, J.}
\newblock \emph{\bibinfo{title}{Deep residual learning for image recognition}}, \bibinfo{pages}{770--778} (\bibinfo{year}{2016}).

\bibitem{hendrycks2016gelu}
\bibinfo{author}{Hendrycks, D.} \& \bibinfo{author}{Gimpel, K.}
\newblock \bibinfo{title}{Gaussian {E}rror {L}inear {U}nits ({GELU}s)}.
\newblock \emph{\bibinfo{journal}{arXiv preprint arXiv:1606.08415}}  (\bibinfo{year}{2016}).

\bibitem{ramachandran2017swish}
\bibinfo{author}{Ramachandran, P.}, \bibinfo{author}{Zoph, B.} \& \bibinfo{author}{Le, Q.~V.}
\newblock \bibinfo{title}{Searching for activation functions}.
\newblock \emph{\bibinfo{journal}{arXiv preprint arXiv:1710.05941}}  (\bibinfo{year}{2017}).

\bibitem{clevert2015fast}
\bibinfo{author}{Clevert, D.-A.}, \bibinfo{author}{Unterthiner, T.} \& \bibinfo{author}{Hochreiter, S.}
\newblock \bibinfo{title}{Fast and accurate deep network learning by exponential linear units ({ELU}s)}.
\newblock \emph{\bibinfo{journal}{arXiv preprint arXiv:1511.07289}}  (\bibinfo{year}{2015}).

\bibitem{misra2020mish}
\bibinfo{author}{Misra, D.}
\newblock \bibinfo{title}{Mish: A self regularized non-monotonic activation function}.
\newblock \emph{\bibinfo{journal}{arXiv preprint arXiv:1908.08681}}  (\bibinfo{year}{2019}).

\bibitem{kordzanganeh2023exponentially}
\bibinfo{author}{Kordzanganeh, M.}, \bibinfo{author}{Sekatski, P.}, \bibinfo{author}{Fedichkin, L.} \& \bibinfo{author}{Melnikov, A.}
\newblock \bibinfo{title}{An exponentially-growing family of universal quantum circuits}.
\newblock \emph{\bibinfo{journal}{Machine Learning: Science and Technology}} \textbf{\bibinfo{volume}{4}}, \bibinfo{pages}{035036} (\bibinfo{year}{2023}).

\bibitem{Schuld2014Quest}
\bibinfo{author}{Schuld, M.}, \bibinfo{author}{Sinayskiy, I.} \& \bibinfo{author}{Petruccione, F.}
\newblock \bibinfo{title}{The quest for a quantum neural network}.
\newblock \emph{\bibinfo{journal}{Quantum Information Processing}} \textbf{\bibinfo{volume}{13}}, \bibinfo{pages}{2567–2586} (\bibinfo{year}{2014}).

\bibitem{Biamonte2017QuantumML}
\bibinfo{author}{Biamonte, J.} \emph{et~al.}
\newblock \bibinfo{title}{Quantum machine learning}.
\newblock \emph{\bibinfo{journal}{Nature}} \textbf{\bibinfo{volume}{549}}, \bibinfo{pages}{195–202} (\bibinfo{year}{2017}).

\bibitem{Cerezo2021VQA}
\bibinfo{author}{Cerezo, M.} \emph{et~al.}
\newblock \bibinfo{title}{Variational quantum algorithms}.
\newblock \emph{\bibinfo{journal}{Nature Reviews Physics}} \textbf{\bibinfo{volume}{3}}, \bibinfo{pages}{625--644} (\bibinfo{year}{2021}).

\bibitem{Cong2019QCNN}
\bibinfo{author}{Cong, I.}, \bibinfo{author}{Choi, S.} \& \bibinfo{author}{Lukin, M.~D.}
\newblock \bibinfo{title}{Quantum convolutional neural networks}.
\newblock \emph{\bibinfo{journal}{Nature Physics}} \textbf{\bibinfo{volume}{15}}, \bibinfo{pages}{1273–1278} (\bibinfo{year}{2019}).

\bibitem{Mitarai2018QCL}
\bibinfo{author}{Mitarai, K.}, \bibinfo{author}{Negoro, M.}, \bibinfo{author}{Kitagawa, M.} \& \bibinfo{author}{Fujii, K.}
\newblock \bibinfo{title}{Quantum circuit learning}.
\newblock \emph{\bibinfo{journal}{Physical Review A}} \textbf{\bibinfo{volume}{98}}, \bibinfo{pages}{032309} (\bibinfo{year}{2018}).

\bibitem{PerezSalinas2020DataReupload}
\bibinfo{author}{Pérez-Salinas, A.}, \bibinfo{author}{Cervera-Lierta, A.}, \bibinfo{author}{Gil-Fuster, E.} \& \bibinfo{author}{Latorre, J.~I.}
\newblock \bibinfo{title}{Data re-uploading for a universal quantum classifier}.
\newblock \emph{\bibinfo{journal}{Quantum}} \textbf{\bibinfo{volume}{4}}, \bibinfo{pages}{226} (\bibinfo{year}{2020}).

\bibitem{Crooks2019ParamShift}
\bibinfo{author}{Crooks, G.~E.}
\newblock \bibinfo{title}{Gradients of parameterized quantum gates using the parameter-shift rule and gate decomposition}.
\newblock \emph{\bibinfo{journal}{arXiv preprint arXiv:1905.13311}}  (\bibinfo{year}{2019}).

\bibitem{Stokes2020QNG}
\bibinfo{author}{Stokes, J.}, \bibinfo{author}{Izaac, J.}, \bibinfo{author}{Killoran, N.} \& \bibinfo{author}{Carleo, G.}
\newblock \bibinfo{title}{Quantum natural gradient}.
\newblock \emph{\bibinfo{journal}{Quantum}} \textbf{\bibinfo{volume}{4}}, \bibinfo{pages}{269} (\bibinfo{year}{2020}).

\bibitem{Sweke2020SGD}
\bibinfo{author}{Sweke, R.} \emph{et~al.}
\newblock \bibinfo{title}{Stochastic gradient descent for hybrid quantum-classical optimization}.
\newblock \emph{\bibinfo{journal}{Quantum}} \textbf{\bibinfo{volume}{4}}, \bibinfo{pages}{314} (\bibinfo{year}{2020}).

\bibitem{Schuld2021Supervised}
\bibinfo{author}{Schuld, M.}
\newblock \bibinfo{title}{{Supervised quantum machine learning models are kernel methods}}.
\newblock \emph{\bibinfo{journal}{arXiv preprint arXiv:2101.11020}}  (\bibinfo{year}{2021}).

\bibitem{Holmes2023Nonlinear}
\bibinfo{author}{Holmes, Z.}, \bibinfo{author}{Coble, N.~J.}, \bibinfo{author}{Sornborger, A.~T.} \& \bibinfo{author}{Suba{\c{s}}{\i}, Y.}
\newblock \bibinfo{title}{Nonlinear transformations in quantum computation}.
\newblock \emph{\bibinfo{journal}{Physical Review Research}} \textbf{\bibinfo{volume}{5}}, \bibinfo{pages}{013105} (\bibinfo{year}{2023}).

\bibitem{DeCross2023QubitReuse}
\bibinfo{author}{DeCross, M.}, \bibinfo{author}{Chertkov, E.}, \bibinfo{author}{Kohagen, M.} \& \bibinfo{author}{Foss-Feig, M.}
\newblock \bibinfo{title}{Qubit-reuse compilation with mid-circuit measurement and reset}.
\newblock \emph{\bibinfo{journal}{Physical Review X}} \textbf{\bibinfo{volume}{13}}, \bibinfo{pages}{041057} (\bibinfo{year}{2023}).

\bibitem{Paetznick2013RUS}
\bibinfo{author}{Paetznick, A.} \& \bibinfo{author}{Svore, K.~M.}
\newblock \bibinfo{title}{Repeat-until-success: Non-deterministic decomposition of single-qubit unitaries}.
\newblock \emph{\bibinfo{journal}{arXiv preprint arXiv:1311.1074}}  (\bibinfo{year}{2013}).

\bibitem{Guerreschi2019RUS}
\bibinfo{author}{Guerreschi, G.~G.}
\newblock \bibinfo{title}{Repeat-until-success circuits with fixed-point oblivious amplitude amplification}.
\newblock \emph{\bibinfo{journal}{Physical Review A}} \textbf{\bibinfo{volume}{99}}, \bibinfo{pages}{022306} (\bibinfo{year}{2019}).

\bibitem{Moreira2023RUS}
\bibinfo{author}{Moreira, M.} \emph{et~al.}
\newblock \bibinfo{title}{Realization of a quantum neural network using repeat-until-success circuits in a superconducting quantum processor}.
\newblock \emph{\bibinfo{journal}{npj Quantum Information}} \textbf{\bibinfo{volume}{9}}, \bibinfo{pages}{118} (\bibinfo{year}{2023}).

\bibitem{Bocharov2015RUS}
\bibinfo{author}{Bocharov, A.}, \bibinfo{author}{Roetteler, M.} \& \bibinfo{author}{Svore, K.~M.}
\newblock \bibinfo{title}{Efficient synthesis of universal repeat-until-success quantum circuits}.
\newblock \emph{\bibinfo{journal}{Physical Review Letters}} \textbf{\bibinfo{volume}{114}}, \bibinfo{pages}{080502} (\bibinfo{year}{2015}).

\bibitem{Wiebe2014RUSCircuit}
\bibinfo{author}{Wiebe, N.} \& \bibinfo{author}{Roetteler, M.}
\newblock \bibinfo{title}{Quantum arithmetic and numerical analysis using repeat-until-success circuits}.
\newblock \emph{\bibinfo{journal}{Quantum Information \& Computation}} \textbf{\bibinfo{volume}{16}}, \bibinfo{pages}{134–178} (\bibinfo{year}{2016}).

\bibitem{Giovannetti2008PRL}
\bibinfo{author}{Giovannetti, V.}, \bibinfo{author}{Lloyd, S.} \& \bibinfo{author}{Maccone, L.}
\newblock \bibinfo{title}{Quantum random access memory}.
\newblock \emph{\bibinfo{journal}{Physical Review Letters}} \textbf{\bibinfo{volume}{100}}, \bibinfo{pages}{160501} (\bibinfo{year}{2008}).

\bibitem{Giovannetti2008PRA}
\bibinfo{author}{Giovannetti, V.}, \bibinfo{author}{Lloyd, S.} \& \bibinfo{author}{Maccone, L.}
\newblock \bibinfo{title}{Architectures for a quantum random access memory}.
\newblock \emph{\bibinfo{journal}{Physical Review A}} \textbf{\bibinfo{volume}{78}}, \bibinfo{pages}{052310} (\bibinfo{year}{2008}).

\bibitem{Rebentrost2014QSVM}
\bibinfo{author}{Rebentrost, P.}, \bibinfo{author}{Mohseni, M.} \& \bibinfo{author}{Lloyd, S.}
\newblock \bibinfo{title}{Quantum support vector machine for big data classification}.
\newblock \emph{\bibinfo{journal}{Physical Review Letters}} \textbf{\bibinfo{volume}{113}}, \bibinfo{pages}{130503} (\bibinfo{year}{2014}).

\bibitem{Yuan2023Optimal}
\bibinfo{author}{Yuan, P.} \& \bibinfo{author}{Zhang, S.}
\newblock \bibinfo{title}{Optimal (controlled) quantum state preparation and improved unitary synthesis by quantum circuits with any number of ancillary qubits}.
\newblock \emph{\bibinfo{journal}{Quantum}} \textbf{\bibinfo{volume}{7}}, \bibinfo{pages}{956} (\bibinfo{year}{2023}).

\bibitem{Breiman1996Bagging}
\bibinfo{author}{Breiman, L.}
\newblock \bibinfo{title}{Bagging predictors}.
\newblock \emph{\bibinfo{journal}{Machine Learning}} \textbf{\bibinfo{volume}{24}}, \bibinfo{pages}{123--140} (\bibinfo{year}{1996}).

\bibitem{Bishop1995Neural}
\bibinfo{author}{Bishop, C.~M.}
\newblock \emph{\bibinfo{title}{Neural Networks for Pattern Recognition}}  (\bibinfo{publisher}{Oxford University Press}, \bibinfo{year}{1995}).

\bibitem{Duan2024Compact}
\bibinfo{author}{Duan, B.} \& \bibinfo{author}{Hsieh, C.-Y.}
\newblock \bibinfo{title}{Compact and classically preprocessed data-loading quantum circuit as a quantum random access memory}.
\newblock \emph{\bibinfo{journal}{Physical Review A}} \textbf{\bibinfo{volume}{110}}, \bibinfo{pages}{012616} (\bibinfo{year}{2024}).

\bibitem{Connor2021ResilienceQRAM}
\bibinfo{author}{Hann, C.~T.}, \bibinfo{author}{Lee, G.}, \bibinfo{author}{Girvin, S.} \& \bibinfo{author}{Jiang, L.}
\newblock \bibinfo{title}{Resilience of quantum random access memory to generic noise}.
\newblock \emph{\bibinfo{journal}{PRX Quantum}} \textbf{\bibinfo{volume}{2}}, \bibinfo{pages}{020311} (\bibinfo{year}{2021}).

\bibitem{kuzmin2025method}
\bibinfo{author}{Kuzmin, V.}, \bibinfo{author}{Somogyi, W.}, \bibinfo{author}{Pankovets, E.} \& \bibinfo{author}{Melnikov, A.}
\newblock \bibinfo{title}{Method for noise-induced regularization in quantum neural networks}.
\newblock \emph{\bibinfo{journal}{Advanced Quantum Technologies}} \textbf{\bibinfo{volume}{8}}, \bibinfo{pages}{e00603} (\bibinfo{year}{2025}).

\end{thebibliography}

\end{document}